\documentclass[aps,prd,twocolumn,showpacs,superscriptaddress,groupedaddress]{revtex4} 
\usepackage[latin1]{inputenc}
\usepackage{graphicx}
\usepackage{amssymb}
\usepackage{color}
\usepackage{float}
\usepackage{amsmath}
\usepackage{amsfonts}
\usepackage{dcolumn}
\usepackage{hyperref}
\usepackage{amsthm}
\usepackage{color}

\def\3nab{\tilde{\nabla}}

\def\be {\begin{equation}}
\def\ee {\end{equation}}
\def\ba {\begin{eqnarray}}
\def\ea {\end{eqnarray}}
\newtheorem{prop}{Proposition}

\newcommand{\barray}{\begin{array}}
\newcommand{\earray}{\end{array}}

\begin{document}
 
\title{Role of equation of states and thermodynamic potentials in avoidance of trapped surfaces in gravitational collapse}
\author{Rituparno Goswami}
 \email{goswami@ukzn.ac.za}
 \affiliation{Astrophysics and Cosmology Research Unit, School of Mathematics, Statistics and Computer Science, University of KwaZulu-Natal, Private Bag X54001, Durban 4000, South Africa.}
 \author{Terricia Govender}
\email{212505648@stu.ukzn.ac.za}
 \affiliation{National Astrophysics and Space Science Program, School of Physics and Chemistry, University of KwaZulu-Natal, Private Bag X54001, Durban 4000, South Africa}

\begin{abstract}
\noindent In this paper we consider the novel scenario where a spherically symmetric perfect fluid star is undergoing continual gravitational collapse while continuously radiating energy in an exterior radiating spacetime. There are no trapped surfaces and the collapse ends to a flat spacetime. Also the collapsing matter obeys the weak and dominant energy conditions at all epoch. Our analysis transparently brings out the role of the equation of state as well as the bounds on the thermodynamic potentials to realise such a scenario. We argue that, since the system of Einstein field equations allows for such a scenario for an open set of initial data as well as the equation of state function in their respective functional spaces, these models are generic and devoid of the problems and paradoxes related to horizons and singularities. The recent high resolution radio telescopes should in principle detect the presence of these compact objects in the sky. 
 \end{abstract}

\pacs{04.20.Cv, 04.20.Dw}

\maketitle


\section{Introduction}
 When a continually collapsing star crosses it's own Schwarzschild radius, it gets trapped. For all the collapsing shells, ingoing as well as outgoing null wavefronts normal to these shells converge and hence the matter must collapse to a central singularity. Existence of these closed trapped 2-surfaces (the collapsing shells after crossing the Schwarzschild radius) is key to all the singularity theorems developed by Hawking, Penrose and Geroch \cite{HE}. The process of formation of trapped surfaces, trapped regions and the boundary of these trapped regions, is also central to the black hole physics. In the context of general relativity, these were first highlighted by Oppenheimer, Snyder and Datt (OSD) \cite{OSD}, for collapse of a pressureless homogeneous dust. It was shown that the entire star gets trapped much before the formation of the central singularity and hence the central singularity can never be seen by far away observers. Although the OSD model is extremely idealised, Penrose argued that any realistic gravitational collapse should be qualitatively similar to this model which became his famous cosmic censorship conjecture \cite{CCC}.\\

This censorship conjecture has no formal mathematical proof/disproof till now. However there are numerous counterexamples for which the idealised picture of the OSD collapse does not hold. There are possibilities of naked singularities that can be visible to faraway observers, before being trapped ( see for example \cite{Joshi} and the references therein). In all these examples the formation of trapped surfaces are delayed by the presence of matter shear or Weyl curvature \cite{AH}, so that part of the singularity becomes naked. But still, in all these counterexamples the trapped surfaces are present in the spacetime. \\

Another novel picture of gravitational collapse of a massive star was first considered by \cite{las}, where a collapsing star is continuously radiating and losing it's mass, and therefore the surface of the star never crosses the Schwarzschild radius. There is no trapped region in the spacetime, and when the shells reach the central point, all the matter is radiated away, leaving a flat spacetime. The exact mechanisms by which the collapsing star can transfer radiation/matter to the external spacetime was studied by numerous authors thereafter. One can match the collapsing star with a non-comoving (evaporating) boundary to a purely radiating Vaidya exterior \cite{sen1,sen2,sen3,sen4,sen5,sen6}. The matching can also be done at a comoving stellar boundary to a generalised Vaidya exterior (see \cite{GOS} and the references therein). The matter form for all the models considered in this scenario are very specific, like a specific kind of self interacting scalar field \cite{GOS2}, or matter fields that have a specific form of negative pressures in the later stages of collapse.\\

In this paper, we carefully investigate the role of the equation of state for the isentropic perfect fluid stellar matter, that can give rise to a collapse without any trapping together with the weak and dominant energy conditions being satisfied. Our analysis brings out transparently the classes of equation of state functions as well as the bounds on the thermodynamic potentials to realise such a scenario. We explicitly relate our results to the enthalpy and acceleration potentials of the collapsing matter.
We show that these classes have non-zero measure in the function space. Therefore, these models are generic in nature and devoid of the problems or paradoxes related to horizons and singularities. The recent high resolution radio telescopes (like Event Horizon Telescope) should in principle detect the presence of these compact objects in the sky as the observational signatures of these continually collapsing but non-trapped compact objects will definitely be different from those of a black hole. \\

The paper is organised as follows: In the next section we describe the field equations along with the regularity and energy conditions for the collapsing matter. Section 3 presents the general solution of the collapsing perfect fluid in terms of the thermodynamic quantities. Section 4 establishes the conditions for no-trapping, with some specific examples. In section 5 we transparently relate our results to the thermodynamic potentials of the collapsing matter. Finally in the last section we give some concluding remarks about the final fate of the models we have considered here.\\

\section{The spherical collapsing star: Field equations and regularity conditions}

\noindent We present a general line element for a spherically symmetric distribution of a collapsing perfect fluid in comoving spherical coordinates $(x^{i}=t,r,\theta,\phi)$ as
\begin{equation} 
ds^2 = -e^{2\nu(t,r)}dt^2 + e^{2\psi(t,r)} dr^2 + R^2(t,r) d\Omega^2.
\end{equation}
For a perfect fluid, the energy-momentum tensor is given by
\begin{equation}
T_{t}^{t}  = -\rho(t,r), T_{r}^{r} = T_{\theta}^{\theta} = T_{\phi}^{\phi} = p(t,r).
\end{equation}

The quantities $\rho$ and $p$ represents the energy density and pressure respectively for the matter cloud. The weak energy conditions are then considered to have been met by the matter fields. Thus, implicating that the energy density (as measured by any local observer) is non-negative and hence for any time-like vector $V^{i}$ we must have
\begin{equation}
T_{ik} V^{i}V^{k} \geq 0,
\end{equation}
which simplifies to $\rho \geq 0$ and $\rho + p \geq 0$. 

We define the Misner-Sharp mass for the collapsing star (which is the mass inside a given comoving radius $r$, at a given time $t$) as,
\begin{equation}
F = 1 - R^{,a}R_{,a} = 1 - G + H, \label{GH}
\end{equation}
where
\begin{eqnarray}
G(t,r) & = & e^{-2\psi}\left(R^\prime\right)^2, \\\label{G}
H(t,r) & = & e^{-2\nu}\left(\dot{R}\right)^2. \label{H}
\end{eqnarray}
\noindent In terms of this mass function the Einstein field equations are given as,
\begin{eqnarray}
F^\prime & = & \rho R^2 R^\prime,\\
\dot{F} & = & - p R^2 \dot{R}, \\
p^\prime & = & -\nu^\prime \left(\rho + p\right), \\ \label{p}
R^\prime \dot{G} & = & 2 \dot{R}\nu^\prime G, \label{0}
\end{eqnarray}

Note that $(\dot{})$ and $({}^\prime)$ depicts partial derivatives with respect to $t$ and $r$ respectively. 
We impose the condition $F(t_i, 0) = 0$ to preserve regularity at the initial epoch. Since we want to investigate the collapsing class of solutions to the Einstein equations, we must impose the condition $\dot{R}<0$ on the area radius $R$. This imposition results in the area radius of all shells of the continual collapsing cloud to monotonically diminish to zero (forming the spacetime singularity), $R(t_s(r),r) = 0$ and where time $t = t_s(r)$ is the time taken for a shell labelled $r$ to reach the singularity. At the  time $t = t_i$, the radius $R = r$, due to the use of the scaling freedom for the radial coordinate $r$. We proceed with an arbitrary scaling function,
\begin{equation}
a(t,r)\equiv \frac{R}{r},
\end{equation}
resulting in,
\begin{eqnarray}
R(t,r) & = & ra(t,r), \label{R}\\
a(t_i,r) & = & 1,\\
a(t_s (r),r) & = & 0,
\end{eqnarray}
where $\dot{a} < 0$. The  regularity conditions suggests that $F \approx r^3$ close to the center. The normal structure of $F$ is thus,
\begin{equation}
F(t,r) = r^3 M(r,a),\label{F}
\end{equation}
where we have $M(r, a)$ to be some general function as restricted by the regularity conditions and energy conditions.

\section{The metric as a function of thermodynamic quantities}

To obtain the general solution of the metric functions in terms of the thermodynamic quantities at any epoch, we perform a change of variables from $(t,r)$ to $(a,r)$. In that case for any function $\Phi(t,r)$, we must have
\begin{eqnarray}
\dot{\Phi}& = &\Phi_{,a}\dot{a}, \\
\Phi^\prime& = &\Phi_{,r} + \Phi_{,a} a_{,r} .
\end{eqnarray}
Also for integrating the $G^0_1=T^0_1$ equation (\ref{0}), we define, 
\begin{equation}
\frac{\nu ^\prime}{R^\prime} \equiv A(r,a)_{,a}, \label{nu}
\end{equation}
where $A(r,a)$ is an arbitrary function of the coordinates $r$ and $a$.
Now we can directly integrate equation (\ref{0}) to obtain
\begin{equation}
G(r,a) = [1+r^2 b_{0}(r)] e^{2rA},
\end{equation}
where $b_{0} (r)$ is a free function of integration.
Therefore from (\ref{G}) we have,
\begin{equation}
e^{2\psi} = \frac{\left(R^ \prime \right)^2}{[1+r^2 b_{0}(r)]e^{2rA}}\,.
\end{equation}

Now we can rewrite the definition of the Misner-Sharp mass in terms of these new variables in the following way:
\be
\sqrt{a}\dot{a}=-e^{\nu}\sqrt{e^{2rA}ab_{0}(r)+ah(r,a)+M(r,a)},\label{sqrt a}
\ee
where we have defined 
\be
h(r,a)  = \frac{e^{2rA}-1}{r^2}.
\ee
Integrating (\ref{sqrt a}) we obtain the equation for the time taken for a shell labelled `$r$', to reach the epoch `$a$' as
\begin{equation}
t(a,r) = \int_{a}^{1} \frac{\sqrt{a}da}{e^\nu \sqrt{e^{2rA}ab_{0}+ah+M}}. \label{time}
\end{equation}
The above is the solution for the scaling function $a(t,r)$ in the integral form.
This immediately gives the singularity curve $t_s(r)$,
which is the collapse end time where the shell labelled `$r$' diminishes to zero area radius ($R=0$),
\begin{equation}
t_{s}(r) = \int_{0}^{1}\frac{\sqrt{a}da}{e^\nu \sqrt{e^{2rA}ab_{0}+ah+M}}. 
\end{equation}

We now have to relate the function $A(r,a)$ to the thermodynamic variables of the collapsing matter. To do so, we first write the $G^0_0=T^0_0$ and  $G^1_1=T^1_1$ equations in terms of the new variables
\begin{eqnarray}
\rho &=& \frac{3M + r[M_{,r} + M_{,a} a^\prime]}{a^2 (a+ra^\prime)},  \label{density eqn}\\
p& =& -\frac{M_{,a}}{a^2}. \label{pressure eqn}
\end{eqnarray}
Substituting $M_{,a}$ into the density equation (\ref{density eqn}) and rearranging, we find, 
\begin{equation}
a^\prime = \frac{3M + rM_{,r}-\rho a^3}{ra^2 (\rho + p)}. \label{aprime1}
\end{equation}
To solve $A$ we first find $a^\prime$ from the definition (\ref{nu}) and (\ref{p}) to obtain,
\begin{eqnarray}
A_{,a} R^\prime &=& -\frac{(p_{,r} + p_{,a} a^\prime)}{p + \rho}, \\
a^\prime &=& \frac{-p_{,r} - (p+\rho)A_{,a}a}{r(p+\rho)A_{,a} + p_{,a}}.\label{aprime2}
\end{eqnarray}
We equate (\ref{aprime1}) and (\ref{aprime2}) to solve for $A_{,a}$,
\begin{equation}
A_{,a}= \frac{-p_{,r} [ra^2 (\rho + p)] - [3M+rM_{,r}-\rho a^3] p_{,a}}{[3M +rM_{,r} -\rho a^3][r(\rho+p)]+[ra^3(\rho+p)^2]},
\end{equation}
where $p_{,r} = c_{s}^2 \rho_{,r}$ and $p_{,a} = c_{s}^2 \rho_{,a}$, where $c_s$ is the speed of sound.
Therefore we can write the function $A(r,a)$ as
\begin{equation}
A =\int_{1}^{a} \frac{-p_{,r} [ra^2 (\rho + p)] - [3M+rM_{,r}-\rho a^3] p_{,a}}{[3M +rM_{,r} -\rho a^3][r(\rho+p)]+[ra^3(\rho+p)^2]} da.\label{the A constraint}
\end{equation}
Now we can immediately write the metric function $e^\nu$ as
\be
e^\nu= \exp\left\{\int_0^r \frac{XY}{Z}\right\} dr
\ee
where $X=-p_{,r} [ra^2 (\rho + p)] - [3M+rM_{,r}-\rho a^3] p_{,a} $, $Y=3M + rM_{,r} + a^3 p$ and finally the last variable is $Z = [ra^2 (\rho +p)^2][3M +rM_{,r}-\rho a^3 + a^3 (\rho+p)]$.
We note that the solution of $M$ depends on the equation of state $p=p(\rho)$. Once the equation of state is supplied, then all the metric functions can be written explicitly in terms of the thermodynamic quantities of the collapsing matter field.

\section{Conditions for no-trapping}

It is well known that for any spherically symmetric spacetimes, a shell labelled `$r$' is trapped, if the Misner Sharp mass enclosed by the shell is greater than the area radius of the shell. Therefore the spherical 2-surface labelled by the co-ordinate `$r$' is trapped if  $F > R$ whereas when $F < R$ the surface is not trapped. It is obvious then, that the boundary of the trapped region or the {\it apparent horizon} is described by the equation
\begin{equation}
 F=R \,.\label{boundary of trapped surface}
\end{equation}
For a continual collapsing matter cloud, regularity conditions imply the avoidance of trapped surfaces at the initial epoch. When the boundary of this cloud is $r = r_b$ and we enforce the condition $M_0(r_b)^2 < 1$. Then the avoidance of trapped surfaces for any shell $r \leq r_b$ occurs at $t=t_i$. If we want to avoid trapping in the complete spacetime we must ensure that througout the spacetime
\be
F<R. \label{no trapping}
\end{equation}
This obviously implies that
\begin{equation}
G-H > 0.\label{GH constraint} 
\end{equation}
Using the definition for the function $H$, we then obtain
\begin{eqnarray}
G - e^{(-2\nu)} (\dot{R})^2 &>& 0, \label{Gee}
\\ \left( \sqrt{G} - e^{(-\nu)} \dot{R} \right) \left( \sqrt{G} + e^{(-\nu)} \dot{R} \right) &>& 0. \label{sqrtG}
\end{eqnarray} 
Since $\dot{R} < 0$ during collapse we then find (\ref{sqrtG}) to be,
\begin{eqnarray}
-e^{(-\nu)} \dot{R}& < &\sqrt{G}.\label{the constraint}
\end{eqnarray}
We can now summarise the conditions for no-trapping in the following way:
\begin{prop}
For continually collapsing spherically symmetric perfect fluid with energy density $\rho(r,a)$ and pressure $p(r,a)$ with $r\in[0,r_b]$ and $a\in[0,1]$, if the following conditions are satisfied:
\begin{enumerate}
\item $\rho>0$ and $\rho+p\ge0$ $\forall$ $r\in[0,r_b]$ and $\forall$ $a\in[0,1]$,
\item $p(r,1)>0$ for all $r\in[0,r_b]$,
\item $r\sqrt{e^{2rA} b_{0} +ah+M} <\sqrt{a(1+r^2 b_{0})}e^{rA}$,  $\forall$ $r\in[0,r_b]$ and $\forall$ $a\in[0,1]$, where the function $A(r,a)$ is given by equation (\ref{the A constraint}),
\end{enumerate}
then the collapsing spacetime will be devoid of any trappings in spite of the weak energy condition being satisfied by the collapsing matter.
\end{prop}
The second condition above, is to ensure that the collapse commences with positive pressure. However in the process of collapse the pressure can be negative without violating the energy conditions. The above conditions get remarkably simple in the case of homogeneous collapse, given by the FLRW metric
\begin{equation}
ds^2 = -dt^2 + a^2 (t) \left(dr^2 + r^2 d\Omega ^2 \right). 
\end{equation}
In this case the equation $G_{0}^{0} = T_{0}^{0}$ can be directly integrated to give
\begin{equation}
M = \frac{1}{3} \rho a^3.\label{misner mass}
\end{equation}
Integrating the energy conservation equation 
\begin{equation}
\dot{\rho} = -3 \frac{\dot{a}}{a} \left(\rho + p(\rho)\right) \label{rho dot},
\end{equation}
gives,
\begin{equation}
a = \exp \left[-\frac{1}{3}\int_{\rho_0}^\rho \frac{d\rho}{\rho + p}\right].\label{a}
\end{equation}
The condition for no trapping is now given by (as $F$ is a monotonic function of $r$ at any given epoch)
\begin{equation}
\frac{{r_b}^3 M}{r_b a} <1.
\end{equation}
Using equations (\ref{misner mass}, \ref{a}) the condition becomes
\begin{equation}
\rho < \frac{3}{{r_b}^2} \exp \left[\frac{2}{3}\int_{\rho_0}^\rho \frac{d\rho}{\rho + p}\right].\label{rho}
\end{equation}
To show that the set of equations of state which ensures such behaviour is indeed non-empty and open set, we explore  the equation of state
 \be
 p(\rho) = p_0 + p_1\rho,
 \ee
 where $p_0$ and $p_1$ are constants.
Then we can directly integrate equation (\ref{rho}) to get the condition
\begin{eqnarray}
 1 &<& \frac{3}{{r_b}^2} \frac{(p_0 + \rho + p_1 \rho)^{ \left(\frac{2}{3(1+p_1)}\right)}}{\rho}.
\end{eqnarray}
We immediately see that for the above condition to be satisfied we must have $p_1<0$, and we can always choose $p_0$ and $p_1$ in such a way that at the initial epoch pressure is positive but at later stages it becomes negative, although the energy conditions are satisfied.

\section{Relating to the thermodynamic potentials}

As defined in \cite{EMM}, for reversible flows of isentropic perfect fluid with the barotropic equation  of state $p=p(\rho)$, the enthalpy $W$ and acceleration potential $\mathcal{A}$, can be defined in the following way:
\ba
 W&=&\exp\left\{\int_{\rho_0}^{\rho} \frac{d\rho}{3(\rho+p)}\right\}\, , \label{W}\\
\mathcal{A}&=& \exp\left\{\int_{p_0}^{p} \frac{dp}{(\rho+p)}\right\}\, . \label{acc}
\ea
These potentials relate directly to the energy and momentum conservation equations respectively. In fact, from equation (\ref{p}) we can immediately see that at a given epoch $t=t_0$, we can write
\be
\mathcal{A}=e^{-\nu}\,.
\ee
This proves that for homogeneous spacetimes $\mathcal{A}=1$ and the dynamics of the spacetime is governed only by the matter enthalpy. In terms of the matter enthalpy the condition for no-trapping for homogeneous matter can be stated in a very compact way:
\begin{prop}
For a collapsing spherically symmetric homogeneous perfect fluid cloud, if the matter energy density is strictly bounded by the matter enthalpy by the following relation
\be
\rho < \frac{3}{{r_b}^2} W^2\,,
\ee
then there will not be any trapped surfaces in the spacetime.
\end{prop}
For inhomogeneous spacetimes the relation is more complicated. However, at any given epoch, we can relate the function $A$ to the acceleration potential as
\be
[\ln \mathcal{A}]'= -A_a\left( a+ \frac{3M + rM_{,r}-\rho a^3}{a^2 (\rho + p)}\right)\;.
\ee
Also, for inhomogeneous spacetimes the matter enthalpy is written in terms of the metric functions as
\be
W=\exp\left\{-\frac13\int_{t_0}^t e^{-\nu}\left(\dot{\psi}+2\frac{\dot{a}}{a}\right)dt\right\}\,.
\ee
Using the above two equations we can implicitly relate these potentials to Proposition 1, to see the role of these potentials in avoidance of trapped surfaces. 

\section{Discussions: The final fate of the collapse}

In the previous two sections we gave the conditions on the equations of state, thermodynamic quantities and potentials that ensures no trapped surfaces in the collapsing perfect fluid spacetime. The next obvious question would be, what will be the final outcome of such a collapse? This question is important, because if the final fate is a strong curvature naked singularity (the singularity has to be naked in the absence of trapping), then these models would definitely be unphysical. The key property of these models are: the collapsing matter cloud continuously throws out radiations and matter in an external radiating spacetime, such that the cloud never crosses it's own Schwarzschild radius. Thus, to determine the final outcome we must take into account the external radiating spacetime. One of the most common spacetimes that can be matched with the interior across a comoving boundary $r=r_b$, 
is the generalised Vaidya spacetime,
\be
ds_{+}^2=-\left(1-\frac{2\mathcal{M}(v,r_v)}{r_v}\right)dv^2-2dvdr_v + r_v^2d\Omega ^2 \label{GV}\,.
\ee
This spacetime describes a combination of {\it Type I} and {\it Type II} matter fields and therefore is ideally suited for our model. 
Matching the first and second fundamental forms across a co-moving matching surface gives
\be
\left[\frac{F}{R}\right]_{int}=\left[ \frac{2\mathcal{M}(v,r_v)}{r_v}\right]_{ext}\,.\label{matching}
\ee

\begin{figure}
	\centering
	\includegraphics[width=0.6\linewidth]{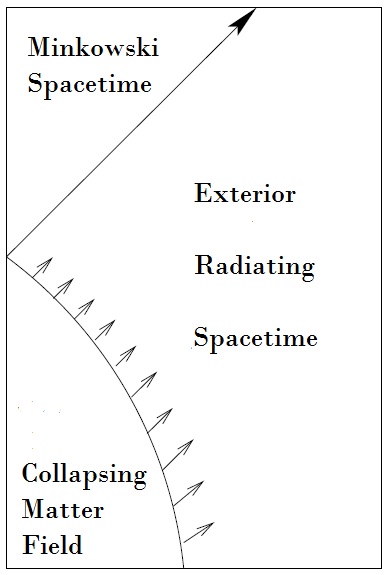}
	\caption{A schematic diagram of the complete spacetime.}
	\label{fig:the-complete-spacetime}
\end{figure}

Now as by construction the LHS of the above equation is restricted to be less than unity, so will the RHS. Therefore for any observer at the exterior spacetime the limit of the generalised Vaidya mass $\mathcal{M}(v,r_v)$, to the  generalised Vaidya radius $r_v$ (at the central singularity $r_v=0$), must be a non-negative number less than unity. This will then give rise to two possible end states:
  \begin{enumerate}
 \item If the limit is non-zero, there will be a naked conical singularity at the centre. These are weak curvature singularities and can be resolved by extending the spacetime through them. 
 \item If the limit goes to zero, then there is no singularity and the collapse ends to a flat spacetime.
 \end{enumerate}
Thus in both cases the collapse ends in a flat spacetime. Fig. 1 describes the schematic diagram of the complete spacetime considered here. An important point to note here is that there exist open sets of equation of state functions in the functional space and also open sets of initial data, for which these models are possible. Therefore the classes of these models have non zero measure in the respective function spaces. Therefore, these models are generic in nature and devoid of the problems and paradoxes related to horizons and singularities. Hence, in principle, these can describe certain kinds of astrophysical collapsing objects, which can be detected by recent high resolution telescopes.

\end{document}